\documentclass[epj]{webofc}
\usepackage[varg]{txfonts}
\usepackage{amsmath} 
\usepackage{bm}
\usepackage{caption}
\usepackage{tikz,graphicx,booktabs,multirow,array,microtype,mathtools,float}
\usepackage[colorlinks=true,backref=false, linktocpage=true,
citecolor=blue,
urlcolor=blue,linkcolor=blue,
pdfpagemode=UseOutlines]{hyperref}
\newcolumntype{C}{>{$}c<{$}}
\usepackage{dcolumn}
\AtBeginDocument{
\heavyrulewidth=.08em
\lightrulewidth=.05em
\cmidrulewidth=.03em
\belowrulesep=.65ex
\belowbottomsep=0pt
\aboverulesep=.4ex
\abovetopsep=0pt
\cmidrulesep=\doublerulesep
\cmidrulekern=.5em
\defaultaddspace=.5em
}

\def\beq{\begin{equation}}
\def\eeq{\end{equation}}
\def\beqs#1\eeqs{\beq\begin{split} #1 \end{split}\eeq}

\def\pmat#1{\begin{pmatrix}#1\end{pmatrix}}

\long\def\comment#1{}

\def\MeV {\mathop{\hbox{MeV}}}

\wocname{EPJ Web of Conferences}
\woctitle{CONF12}
\begin{document}
\selectlanguage{english}
\title{Role of the strange quark in the $\rho(770)$ meson}
%
%

\author{R. Molina\inst{1}\fnsep\thanks{\email{ramope71@email.gwu.edu}} \and
       D. Guo\inst{1} \and
      B. Hu \inst{1}
    \and A. Alexandru\inst{1}\and
    M. Doering\inst{1,2}
}

\institute{Physics Department, The George Washington University, Washington, DC 20052, USA
\and
          {Thomas Jefferson National Accelerator Facility, Newport News, VA
23606, USA}
}

\abstract{%
 Recently, the GWU lattice group has evaluated high-precision phase-shift data for $\pi\pi$ scattering
in the $I = 1$, $J = 1$ channel. Unitary Chiral Perturbation Theory describes these data well around
the resonance region and for different pion masses. Moreover, it allows to extrapolate to the physical
point and estimate the effect of the missing $K\bar{K}$ channel in the two-flavor lattice calculation. The
absence of the strange quark in the lattice data leads to a lower $\rho$ mass, and the analysis with
U$\chi$PT shows that the $K \bar{K}$ channel indeed pushes the $\pi\pi$-scattering phase shift upward, having
a surprisingly large effect on the $\rho$-mass. The inelasticity is shown to be compatible with the
experimental data. The analysis is then extended to all available two-flavor lattice simulations and
similar mass shifts are observed. Chiral extrapolations of $N_f = 2 + 1$ lattice simulations for the
$\rho(770)$ are also reported.
}
\maketitle
\section{Introduction}
\label{intro}
\begin{figure}[t]
\begin{center}
\begin{tabular}{cc}
\includegraphics[width=0.5\textwidth]{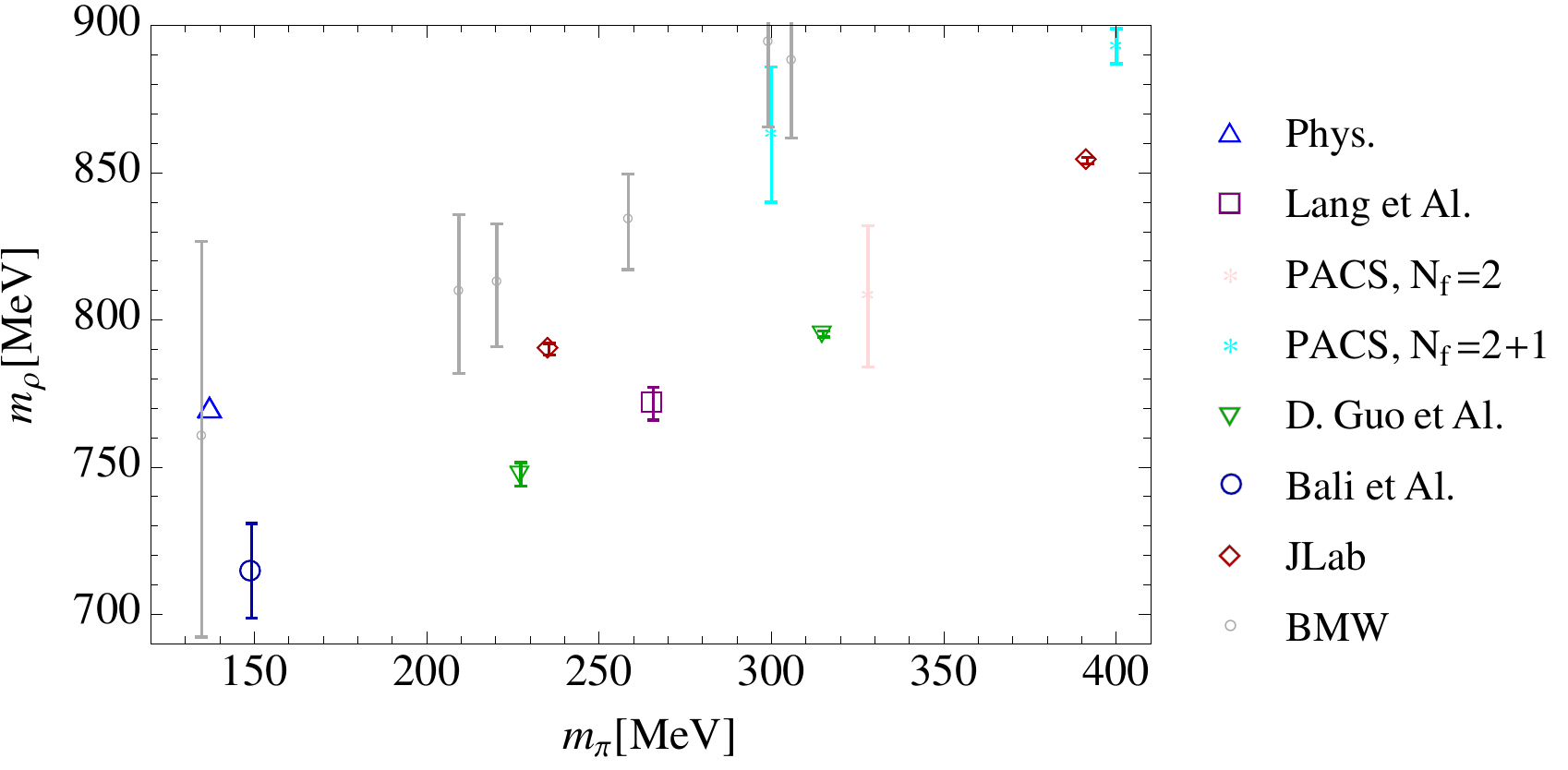}&
\includegraphics[width=0.45\textwidth]{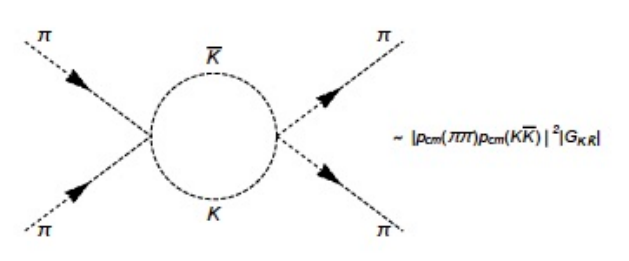}
\end{tabular}
\end{center}
\caption{Left: Mass of the $\rho$-resonance extracted from different simulations, in $N_f=2$ simulations, from Refs. \cite{Bali:2015gji,Guo:2016zos,Lang:2011mn,Aoki:2007rd} and, for $N_f=2+1$, see Refs. \cite{Aoki:2011yj,Feng:2010es,Dudek:2012xn,Wilson:2015dqa}. Right: Insertion of a $K\bar K$ intermediate state in $\pi\pi$
scattering.}
\label{fig:latticedata}
\end{figure}

Recently, extraordinary progress has been achieved in lattice-QCD simulations. Concerning the $\rho$ resonance, Bali {\it et al.} \cite{Bali:2015gji} extracted the $\rho$-resonance parameters from a lattice-QCD simulation for a pion mass very close to the physical one, $m_\pi=149.5$ MeV.
Unexpectedly, the $\rho$-mass extracted in the $N_f=2$ simulation falls below the experimental $\rho$-mass by around $50$ MeV.
High-precision simulations from the GWU \cite{Guo:2016zos} and JLab \cite{Dudek:2012xn,Wilson:2015dqa} groups for a pion mass$\sim 230$ MeV, in $N_f=2$ and $N_f=2+1$, respectively, report values of the $\rho$-mass which are in disagreement between each other.
Generally speaking, $N_f=2$ simulations understimate the extracted values of $\rho$-mass  in $N_f=2+1$ simulations, taking into consideration error bars, as it is shown in Fig. \ref{fig:latticedata}.
In this talk, we analyze the source of differences between several lattice-QCD simulations for the $\pi\pi$ scattering in the $I=1$, $J=1$ channel in the elastic region. The model used to analyze the lattice-QCD data is Unitary Chiral Perturbation Theory adapted to the conditions of the finite volume.

In particular, we estimate the effect of the coupling of the $\rho$ resonance to the $K\bar{K}$ channel. Because of the small observed inelasticity in the $\rho$ channel~\cite{Protopopescu:1973sh} and the small $K\bar K$ phase shifts
obtained in analyses~\cite{Oller:1998hw, Doring:2013wka, Wilson:2015dqa}, it has
been assumed that the $\rho$ meson effectively decouples from the $K\bar K$
channel. Nevertheless, consider an intermediate $K\bar{K}$ loop in the transition $\pi\pi\to\pi\pi$, as depicted in Fig. \ref{fig:latticedata} (right). Since we are dealing with
$p$-waves, the behaviour close to the thresolds is $|p_{\rm cm}(\pi\pi)p_{\rm cm}(K\bar K)|^2|G_{K\bar{K}}|$, being $p_{\rm cm}$ the momenta in the center of mass. This function is zero in the 
$\pi\pi$ and $K\bar{K}$ threshols and shows a maximum around the $\rho$ mass. The above expression has to be multiplied by an unknown function whose form will depend on the $\pi\pi\to K\bar{K}$ transition but that essentially varies smoothly with energy. Moreover, the ratio of the couplings of the $\rho$ meson to $K\bar{K}$ and $\pi\pi$ has been calculated
in NLO and one-loop NLO U$\chi$PT in Refs. \cite{Oller:1998hw,Guo:2011pa}. Both calculations give a value $g_{K\bar{K}}/g_{\pi\pi}\simeq 0.5-0.6$, which is not negligible.

In the present talk, first, the GWU $N_f=2$ lattice eigenvalues are fitted to the U$\chi$PT model. Then, we show the extrapolation to the physical point and an estimation of the $2\to 3$ flavor extrapolation including the $K\bar{K}$ channel. The result is compared with other available $N_f=2$ and $N_f=2+1$ lattice data. Finally, other $N_f=2$ lattice data are fitted using the U$\chi$PT model, and the results for $\rho$ mass and width obtained are compared between the different lattice groups.
\section{Unitarized chiral perturbation theory model}

Chiral Perturbation Theory ($\chi$PT) is an effective field theory of QCD which describes successfully the meson-meson interaction at low energies~\cite{gasser1,gasser2}. 
However, the perturbative expansion is only valid below the energy region where resonances, as the
$\sigma$ and $\rho$ meson, show up. Unitarity in coupled-channels allows to extend the theory to higher energies, and can constraint the pole positions of these resonances and the low-energy $\pi\pi$ scattering amplitude \cite{Colangelo:2001df,Pelaez:2015qba,Oller:1998hw}.
Unitarized Chiral Perturbation Theory is thus a nonperturbative method which combines constraints from chiral symmetry and its breaking and 
(coupled-channel) unitarity. The method of Ref.~\cite{Oller:1998hw} is able to describe 
the meson-meson interaction up to about $1.2$ GeV. The scattering amplitudes develop
poles in the complex plane which can be associated with the known scalar and vector resonances. 
The Inverse Amplitude method relies upon an expansion of the inverse of the $T$-matrix in powers of the momenta, which has better convergence around the resonances \cite{Truong1}. The $T$-matrix can be written as 
~\cite{Oller:1998hw}
\begin{equation}
T=[I-VG]^{-1}V
\label{eq:tmat2}
\end{equation}
where
\begin{equation}
V=V_2[V_2-V_4]^{-1}V_2\,.
\label{eq:viam}
\end{equation}
In the above equation, $V_2$ and $V_4$ are the respective potentials evaluated from the ${\cal O}(p^2)$ and ${\cal O}(p^4)$ chiral Lagrangians \cite{gasser1,gasser2}. In Eq.
 (\ref{eq:tmat2}), $G$ is a diagonal matrix whose elements are the two-meson loop functions, 
evaluated in our case in dimensional regularization in contrast to the cutoff-scheme used in the original model of Ref.~\cite{Oller:1998hw}:
\begin{eqnarray}
G^{DR}_{ii}(E)
&&= i \,  \int \frac{d^4 q}{(2 \pi)^4} \,
\frac{1}{q^2 - m_{1}^2 + i \epsilon} \, \frac{1}{(P-q)^2 - m_{2}^2 + i
\epsilon}\label{eq:floop}\\
 &&= \frac{1}{16 \pi^2} \left\{ a(\mu) + \ln
\frac{m_{1}^2}{\mu^2} \right.\left.+\frac{m_{2}^2-m_{1}^2 + E^{2}}{2E^{2}} \ln \frac{m_{2}^2}{m_{1}^2}   +
\frac{p_i}{E}
\left[
\ln(\hspace*{0.2cm}E^{2}-(m_{1}^2-m_{2}^2)+2 p_i E)
\right. \right.
 \nonumber\\&&+ \ln(\hspace*{0.2cm}E^{2}+(m_{1}^2-m_{2}^2)+2 p_i E)-\ln(-E^{2}+(m_{1}^2-m_{2}^2)+2p_i E)  \nonumber\\&&\left. \left.- \ln(-E^{2}-(m_{1}^2-m_{2}^2)+2 p_i E) \right]
\right\} ,
\label{eq:gdimre}
\end{eqnarray}
 where $p_i=\frac{\sqrt{(E^2-(m_1+m_2)^2)(E^2-(m_1-m_2)^2)}}{2 E}$ for the channel $i$, $E$ is the center-of-mass energy, and $m_{1,2}$ refers to the masses of the mesons $1,2$ in the $i$ channel. Throughout this study we use $\mu=1$~GeV and a natural value of the subtraction constant $\alpha(\mu)=-1.28$.
 

The potential of Eq. (\ref{eq:viam}), after projecting in $I=1$, $L=1$, 
only depends on two parameters ~\cite{Oller:1998hw}, and reads
\begin{equation}
 V(\pi\pi)=\frac{-2\,p^2}{3(f_\pi^2-8\,\hat{l}_1 m_\pi^2+4\, \hat{l}_2 E^2)}\ .\label{eq:vpipi}
\end{equation}
In the above equation, specific combinations of the LECs in Ref. ~\cite{Oller:1998hw} have been introduced, $\hat{l}_1\equiv 2\,L_4+L_5$ and $\hat{l}_2\equiv2\,L_1-L_2+L_3$, which are not identical to the SU(2) CHPT low-energy constants. 
In the present study, we use the one-channel $\pi\pi$ potential of Eq. (\ref{eq:vpipi}), which contains the lowest- and next-to-leading-order contact-term contributions, to fit the $N_f=2$ lattice data.

\subsection{Coupled channel case ($\pi\pi-K\bar K$) }\label{sec:cop}

In this section we address the two-coupled-channel case. The interaction in the $(\pi\pi,K\bar{K})$ system, is evaluated from the ${\cal O}(p^2)$ and ${\cal O}(p^4)$ 
Lagrangians of the $\chi$PT expansion~\cite{gasser1,gasser2}. The potentials, $V_2$ and $V_4$, 
projected in $I=1$ and $L=1$ are~\cite{Oller:1998hw}
\beq
V_{2}(E)=-
\pmat{\frac{2p^2_\pi}{3f_\pi^2} & \frac{\sqrt{2}p_K p_\pi}{3 f_Kf_\pi}\\
\frac{\sqrt{2}p_K p_\pi}{3 f_Kf_\pi} & \frac{p_K^2}{3 f_K^2}}
\label{eq:v2c}
\eeq
and 
\beqs
V_{4}(E)=-\pmat{
\frac{8p_\pi^2(2\hat{l}_1 m_\pi^2-\hat{l}_2 E^2)}{3f_\pi^4} & \frac{8p_\pi p_K(L_5(m_K^2+m_\pi^2)-L_3 E^2)}{3\sqrt{2}f_\pi^2 f_K^2} \\
\frac{8p_\pi p_K(L_5(m_K^2+m_\pi^2)-L_3 E^2)}{3\sqrt{2}f_\pi^2 f_K^2} & \frac{4p^2_K (10\hat{l}_1 m_K^2+3(L_3-2\hat{l}_2)E^2)}{9f_K^4}}
\,.
\label{eq:v4c}
\eeqs

Note that the potentials in Eqs.~(\ref{eq:v2c}) and (\ref{eq:v4c}) depend on four low energy constants, 
$\hat{l}_1$, $\hat{l}_2$, $L_3$ and $L_5$. Specifically, the $\hat{l}_1$ and $\hat{l}_2$ parameters control the diagonal transitions, $\pi\pi\to\pi\pi$ and $K\bar{K}\to K\bar{K}$, while the off-diagonal elements $\pi\pi\to K\bar{K}$ are restricted by $L_3$ and $L_5$. When the $N_f=2$ lattice data are fitted, Eq. (\ref{eq:vpipi}) is used. The values of $\hat{l}_1$ and
$\hat{l}_2$ obtained there are used in Eqs.~(\ref{eq:v2c}) and (\ref{eq:v4c}) to extrapolate from the $N_f=2$ to the $N_f=2+1$ case. The other two LECs, $L_3$ and $L_5$, are taken from a global fit to experimental $\pi\pi$ and $\pi K$ phase shifts similarly as in Ref.~\cite{Doring:2013wka}.

In general, the partial wave decomposition of the scattering amplitude of two spinless mesons with definite 
isospin $I$ can be written as 
\begin{equation}
T_I=\sum_J(2 J+1)T_{IJ}P_J(\cos\theta)\,.
\end{equation}
where
\begin{equation}
T_{IJ}=\frac{1}{2}\int^{1}_{-1} P_J(\cos\,\theta)T_I(\theta)\,\,{\rm d}\cos\theta\,.
\end{equation}
Omitting the $I,J$ labels from here on, the two-channel $T$-matrix is evaluated as \cite{Oller:1998hw},
\begin{equation}
 T=V_2[V_2-V_4-V_2GV_2]^{-1}V_2
\end{equation}
In the case of two coupled channels, $T$ $ (\equiv T_{IJ})$ is a $2\times 2$ matrix whose elements $(T)_{ij}$ are
related to $S$ matrix elements through the equations 
\beqs
&(T)_{11}=-\frac{8\pi E}{2ip_1}[(S)_{11}-1]\ ,\quad (T)_{22}=-\frac{8\pi E}{2ip_2}[(S)_{22}-1]\,,
\\
&(T)_{12}=(T)_{21}=-\frac{8\pi E}{2i\sqrt{p_1p_2}}(S)_{12}\,,
\label{eq:sma2}
\eeqs
with $p_1$, $p_2$ the center-of-mas momenta of the mesons in channel $1$ ($\pi\pi$) or $2$ ($K\bar K$) respectively, that is
$p_i=\sqrt{(E/2)^2-m_i^2}$. Finally, the $S$-matrix can be parametrized as
\begin{equation}
S=\pmat{\eta e^{2i\delta_1}& i(1-\eta^2)^{1/2}e^{i(\delta_1+\delta_2)}\\
i(1-\eta^2)^{1/2}e^{i(\delta_1+\delta_2)}& \eta e^{2i\delta_2}}\,.
\label{eq:sma}
\end{equation}

\subsection{Meson-meson scattering in the finite volume and U$\chi$PT model}
For the two-pion system in a finite box, only discrete momenta are allowed, such that, for an asymmetric box with elongation $\eta$ in the $z$ direction , we have
\begin{equation}
 \bm q=\frac{2\pi}{L}(n_x,n_y,n_z/\eta)\ ,
\end{equation}
and the two-meson function loop can be evaluated replacing the integral in Eq. (\ref{eq:floop}) by a sum over the momenta, $\tilde{G}$,
\begin{equation}
\tilde{G}(E)=\frac{1}{\eta L^3} \sum_{\bm q} I(E, \bm q)\,,
\label{eq:gbox}
\end{equation}
where the channel index has been omitted. 
The sum over the momenta is cut off at $q_\text{max}$. Here,
\begin{equation}
I(E, {\bf q})= \frac{\omega_1(\bm q) + \omega_2(\bm q)}
{2 \omega_1(\bm q) \omega_2(\bm q)}\frac{1}{E^2-(\omega_1(\bm q)^2 + \omega_2(\bm q)^2)}\,,
\label{eq:iq}
\end{equation}
 The formalism can also be made independent of 
$q_\text{max}$ and related to the subtraction constant in the dimensional-regularization method, 
$\alpha$ (as in the continuum limit), see Ref~\cite{MartinezTorres:2011pr}, 
\beqs
\tilde{G}=G^{DR}+\lim_{q_\text{max} \to \infty} 
\left( \frac{1}{\eta L^3}\sum_{q<q_\text{max}} I(E,\bm q) - \int_{q<q_\text{max}} 
\frac{d^3 q}{(2\pi)^3} I(E,\bm q) \right) 
\equiv G^{DR} + \lim_{q_\text{max} \rightarrow \infty} \delta G\,,
\label{eq:gtil}
\eeqs
where $G^{DR}$ stands for the two-meson loop function given in Eq.~(\ref{eq:gdimre}).
The scattering amplitude in the finite volume is evaluated similarly as in Eq. (\ref{eq:tmat2}),
\begin{equation}
\tilde{T}=[I-V\tilde{G}]^{-1}V\,,
\label{eq:tmatf}
\end{equation}
or $\tilde{T}=[V^{-1}-\tilde{G}]^{-1}$. Therefore, the energy spectrum in the finite volume can be identified with the poles of the $\tilde{T}$ scattering amplitude, which satisfy the condition,
\begin{equation}
\mathrm{Det}[V^{-1}(E)-\tilde{G}(E)]=0 \ .
\end{equation}
In the one channel case, this corresponds to the energies given by $\tilde{G}(E)=V^{-1}$. Hence, the amplitude in the infinite volume can be evaluated for these energies,
\begin{equation}
T=[\tilde{G}(E)-G(E)]^{-1}\ ,
\label{eq:difg}
\end{equation}
which is independent of the renormalization of the individually divergent expressions. In the case of the two-pion system interacting in $p$-wave, and moving with $\bm{P}=\frac{2\pi}{\eta L}(0,0,1)$ in the direction
of the elongation of the box, the following relations are found,
\begin{eqnarray}
 \hspace{3cm}&&A_2^-:\qquad-1+V(\pi\pi)\tilde{G}_{10,10}=0\label{eq:sym1}\\
 \hspace{3cm}&&E^-:\qquad-1+V(\pi\pi)\tilde{G}_{11,11}=0\ ,\label{eq:sym2}
\end{eqnarray}
with
$\tilde{G}_{lm,l'm'}$ given in Ref.~\cite{Doring:2012eu} but modified as in Eqs.~(\ref{eq:gbox}) and (\ref{eq:gtil}) by the elongation factor $\eta$. $V(\pi\pi)$ is given by
Eq.~(\ref{eq:vpipi}). The above relations are used to fit the energy levels extracted from the lattice.
In Refs.~\cite{Doring:2011vk,Doring:2011ip,Doring:2009bi}, a more detailed presentation of the formalism is presented. In particular, in Ref. \cite{Doring:2011vk}, it is shown that this formalism is equivalent to the L\"uscher approach up to contributions that are exponentially suppressed with the volume. See also Ref.~\cite{Doring:2012eu} for the generalization to the cases of moving frame and partial-wave-mixing in coupled channels. In the present study, $F$-waves were neglected. 
 
\section{Results}
\begin{figure}[b]
\centering
\includegraphics[scale=0.75]{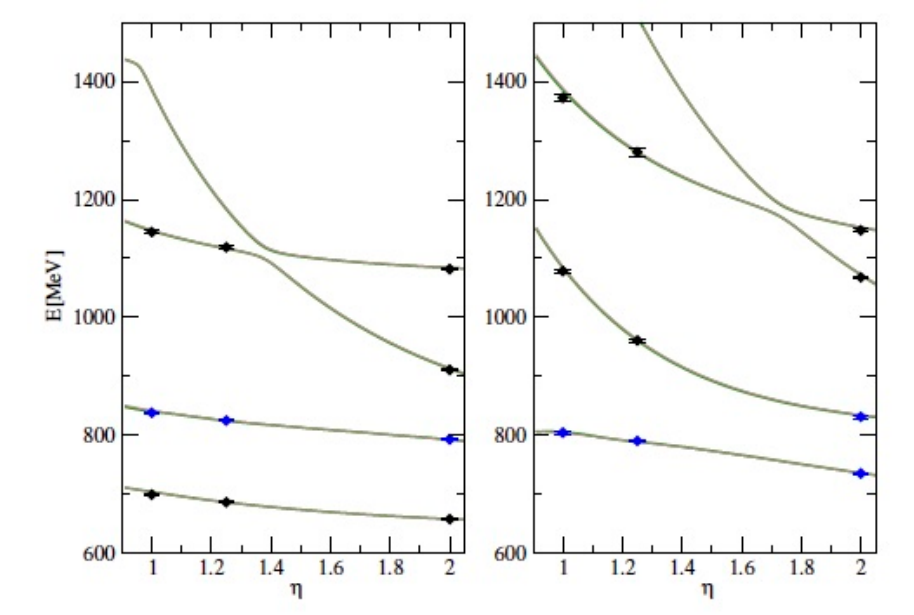}
\caption{Energy spectrum as a function of the elongation factor from unitary chiral perturbation theory for a boost $\mathbf{P}=(0,0,1) 2\pi/\eta L$ (left), and in the rest frame $\mathbf{P}=(0,0,0)$~(right). Dots in blue color fall in the energy region $m_\rho\pm 2\Gamma_\rho$.}
\label{Fig:expectation}
\end{figure}

The energy levels extracted from the GWU lattice simulation in Ref. \cite{Guo:2016zos} are fitted to the U$\chi$PT model in the finite volume using Eqs. (\ref{eq:vpipi}) and (\ref{eq:sym1}). The energy spectrum obtained for a boost $\bm{P}=(0,0,1) 2\pi/\eta L$, and in the rest frame, $\bm{P}=(0,0,0)$, together with the lattice data for $m_\pi=315$ MeV, are shown in Fig. \ref{Fig:expectation}, left and right, respectively. As can be seen in this figure, the U$\chi$PT model describes quite well the lattice data.
 The input required by the model is the pion mass, the pion decay constant, the energy levels and covariance matrices, which are given in Ref. \cite{Guo:2016zos}.

Generally speaking, the U$\chi$PT is able to capture the broad features of the phase shift in the elastic region, however, the very precise determination of the 
energy levels evaluated by the GWU constraints sufficiently the phase shift so that the quality of the fit is not good when trying to fit in the entire energy range. Since we are interested in the mass and width of the $\rho$ resonance, which are well determined by the data in the central energy region, we restrict the fit to the range $m_\rho\pm 2\Gamma_\rho$.

The $\hat{l}_1$ and $\hat{l}_2$ obtained in separate fits for the light and heavy masses are given in Table \ref{table:uchiptfit1}. The resonance mass is determined
from the center-of-mass energy that corresponds to a $90^\circ$ phase-shift. The
width is given by twice the imaginary part of the resonance pole position in the complex
plane. The mass and width obtained are consistent with the ones
determined from a Breit-Wigner fit \cite{Guo:2016zos}. From Table \ref{table:uchiptfit1}, we see that the values of $\hat{l}_1$ and $\hat{l}_2$ for the 
fits to the different pion masses are consistent with each other, what means that the quark mass dependence in the phase shift is well captured. This allows us to combine both results by performing a global fit of both lattice data for light and heavy pion masses. In this case, the quality of the fit is similar to the individual ones, as shown in Table \ref{table:uchiptfit1}, but the $\hat{l}_1$ and $\hat{l}_2$ are determined with better precision.

\begin{table}[t]
\caption[left]{U$\chi$PT fits in the 
$m_\rho\pm 2\Gamma_\rho$ region and extrapolations to the physical point.
The errors quoted are statistical. The upper two entries show the cases of heavy and light pion mass, 
both individually extrapolated to the physical point. 
The third entry shows the combined fit of both masses and its extrapolation.}
\label{table:uchiptfit1}
\vspace{-0.2cm}
\begin{tabular*}{0.99\columnwidth}{@{\extracolsep{\stretch{1}}}c*{5}{r}@{}}
\toprule 
$m_\pi\,[\MeV]$& $\hat l_1\times 10^3$ & $\hat l_2\times 10^3$ & $m_\rho[\MeV]$ & $\Gamma_\rho[\MeV]$  & $\chi^2/\text{dof}$\\ 
\midrule
315 & 1.5(5) & -3.7(2) & 796(1) & 35(1) & 1 \\
138 &           &            & 704(5) & 110(3)& \\
\cmidrule{1-6}
226 & 2(1) & -3.5(2) & 748(1) & 77(1) & 1.53 \\
138 &          &            & 719(4) & 120(3) &  \\
\cmidrule{1-6}
combined & 2.26(14) & -3.44(3) &  &  & 1.26 \\
138 &          &            & 720(1) & 120.8(8) &  \\
\bottomrule
\end{tabular*}
\end{table}

Moreover, since the lattice data do not contain the strange quark, we can estimate the effect of allowing the $\pi\pi$ channel couple to $K\bar{K}$, as described in Section \ref{sec:cop}. The estimates for the $\rho$ mass and width, $\hat{m}_\rho$ and $\hat{\Gamma}_\rho$, are given in Table \ref{table:uchiptfit2} in comparison to the results from the combined fit in the one-channel case.

\begin{table}[b]
\caption[left]{U$\chi$PT results for $N_f=2$, $m_\rho$ and $\Gamma_\rho$, and $N_f=2+1$ estimates,
$\hat m_\rho$ and $\hat \Gamma_\rho$.
The parameters $\hat l_{1,2}$ are taken from the combined fit and the K$\overline{\rm K}$ channel
parameters are taken from fits to experimental data.
The first set of errors quoted are statistical; for $\hat m_\rho$ and $\hat\Gamma_\rho$ we also quote
a set of systematic errors associated with model dependence.}
\label{table:uchiptfit2}\vspace{-0.3cm}
\begin{center}
\begin{tabular*}{0.99\columnwidth}{@{\extracolsep{\stretch{1}}}c*{5}{r}@{}}
\toprule 
$m_\pi\,[\MeV]$& $m_\rho[\MeV]$ & $\Gamma_\rho[\MeV]$  & $\hat m_\rho[\MeV]$ & $\hat \Gamma_\rho[\MeV]$  &\\ 
\midrule
315 & 795.2(7) & 36.5(2) & 846(0.3)(10) & 54(0.1)(3) \\
226 & 747.6(6) & 77.5(5) & 793(0.4)(10) & 99(0.3)(3) \\
138 & 720(1) & 120.8(8) & 766(0.7)(11) & 150(0.4)(5) \\
\bottomrule
\end{tabular*}
\end{center}
\end{table}

\begin{figure}[t]
\begin{center}
\includegraphics[width=0.5\columnwidth]{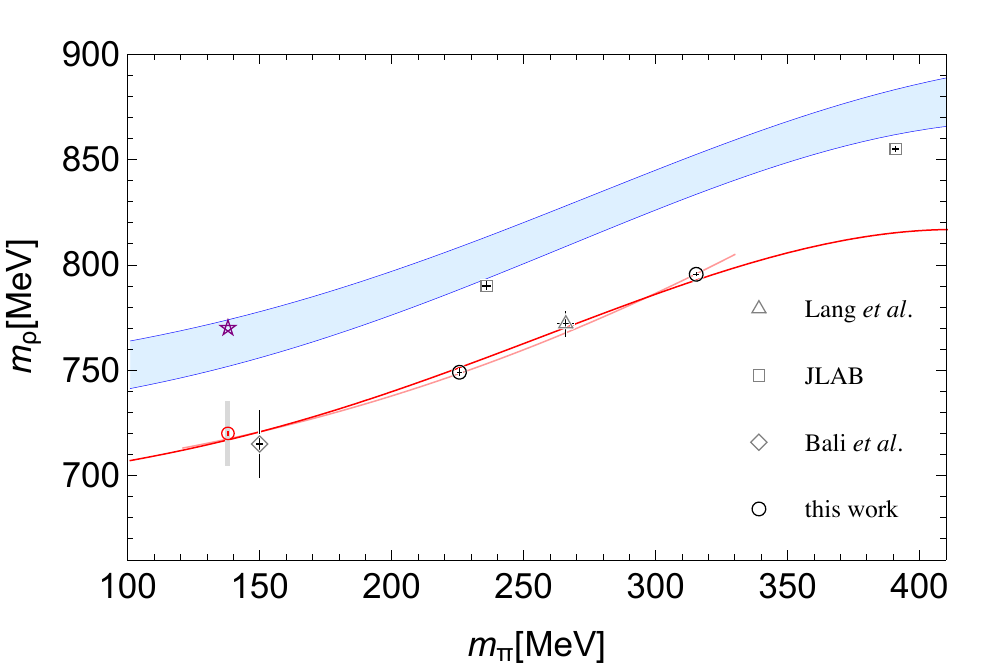}
\end{center}
\caption{Resonance mass extrapolation to the physical point. The red curve corresponds
to an extrapolation based on the U$\chi$PT model. The light-red curve corresponds to 
a simple $m_\rho = (m_\rho)_0 + \text{const}\times m_\pi^2$ fit~\cite{Bruns:2004tj}. The blue band
corresponds to an $N_f=2+1$ estimate based on the U$\chi$PT model (see text). The other
lattice data-points are taken from Lang {\it et al}~\cite{Lang:2011mn},
JLab group studies~\cite{Dudek:2012xn,Wilson:2015dqa}, and Bali~{\it et al}~\cite{Bali:2015gji}.
The star corresponds to the physical result. The error-bars shown with solid lines are stochastic.
For the extrapolation the gray, thick error-bar indicates the systematic error associated with
the lattice spacing determination.}
\label{fig:extrapolation}
\end{figure}

The mass of the resonance as a function of the pion mass obtained from the fit to GWU lattice is shown in Fig. \ref{fig:extrapolation} in comparison with other lattice simulations in $N_f=2$ and $N_f=2+1$. Clearly, the extrapolation to the physical point in SU(2) is significantly lower, around $50$ MeV below the physical mass. The shift cannot be accommodated by the errors in the lattice simulation, even if the systematic errors due to the lattice spacing determination
 are considered. Furthermore, the results from the GWU group are in line with those obtained by Lang {\it et al.} ~\cite{Lang:2011mn} and Bali {\it et al.} ~\cite{Bali:2015gji}. The curve obtained by fitting the GWU's lattice data to the U$\chi$PT model describes quite well the tendency of the $N_f=2$ lattice data.

When estimating the mass of the $\rho$ resonance including the effect of the $K\bar{K}$ channel, the $\rho$ mass is shifted appreciably, and the estimated curve $m_\rho(m_\pi^2)$ with error band is plotted in blue color in Fig. \ref{fig:extrapolation}. The mass of the resonance agrees quite well with the $N_f=2+1$ lattice calculations by the JLab group in Refs.~\cite{Dudek:2012xn,Wilson:2015dqa}, and with the physical mass. Therefore, we conclude that the 
discrepancies between the $N_f=2$ and $N_f=2+1$ lattice data are mostly due to the absence of the strange quark in $N_f=2$ simulations.

\begin{figure}[b]
\begin{center}
\includegraphics[width=0.5\columnwidth]{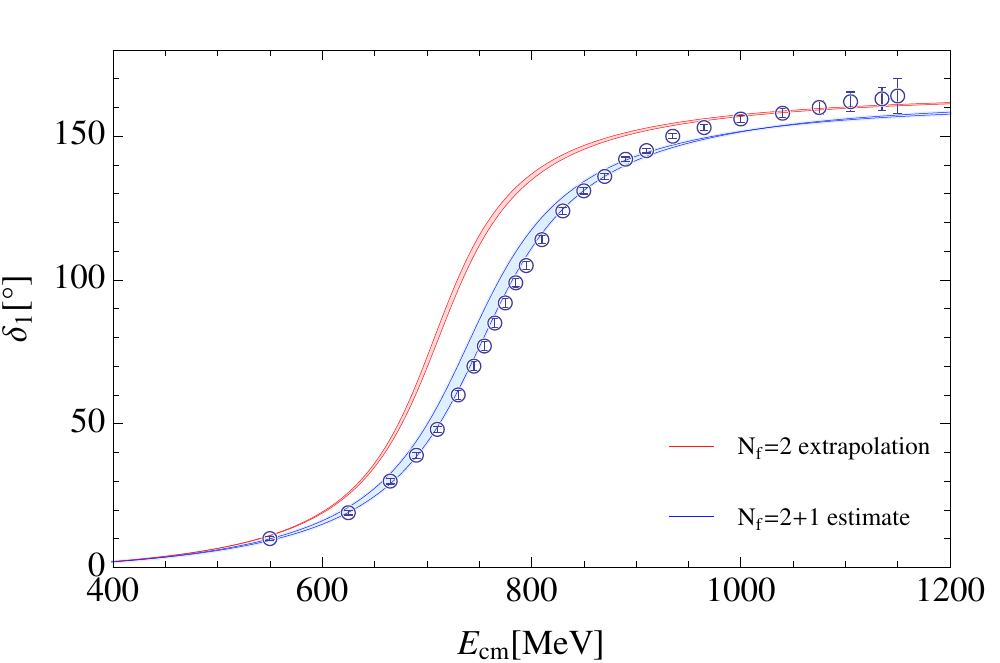}
\end{center}
\caption{Chiral extrapolation of the phase shift to
the physical mass (red band), obtained from the simultaneous fit to lattice eigenvalues at both considered pion masses. Only statistical uncertainties are indicated. The blue band
shows the estimated phase shift when including also the $K\bar{K}$ channel in the two variants mentioned in the text (estimate of systematic uncertainties). To keep the figure simple, statistical uncertainties are not indicated for these cases. They are of the same size as the red band. 
Open circles indicate phase shifts extracted from experiment~\cite{Protopopescu:1973sh}.}
\label{fig:fitfincb}
\end{figure}

The phase shift obtained in the $2\to 3$ flavor extrapolation is shown in Fig. \ref{fig:fitfincb}. The error band calculated in Fig. \ref{fig:extrapolation}(\ref{fig:fitfincb}), blue band, is evaluated by simply letting the $\hat{l}_1$ and $\hat{l}_2$ in the $K\bar{K}\to K\bar{K}$ channel being the ones obtained
in a fit to experimental data, upper(right) curve, or fixed by the lattice data, bottom(left) curve. The difference between both curves is $20$ MeV. If taking the central value, this corresponds to a systematic error of $10$ MeV in the flavor extrapolation.

\begin{figure*}[t]
\begin{center}
\begin{tabular}{ll}
\includegraphics[width=0.45\textwidth]{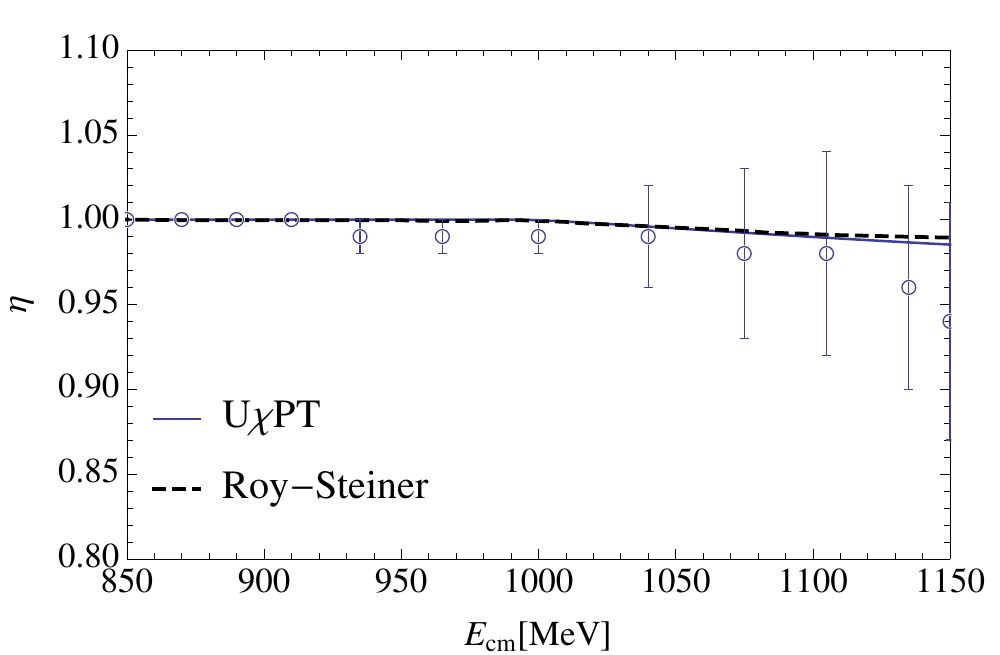}\kern5mm
\includegraphics[width=0.45\textwidth]{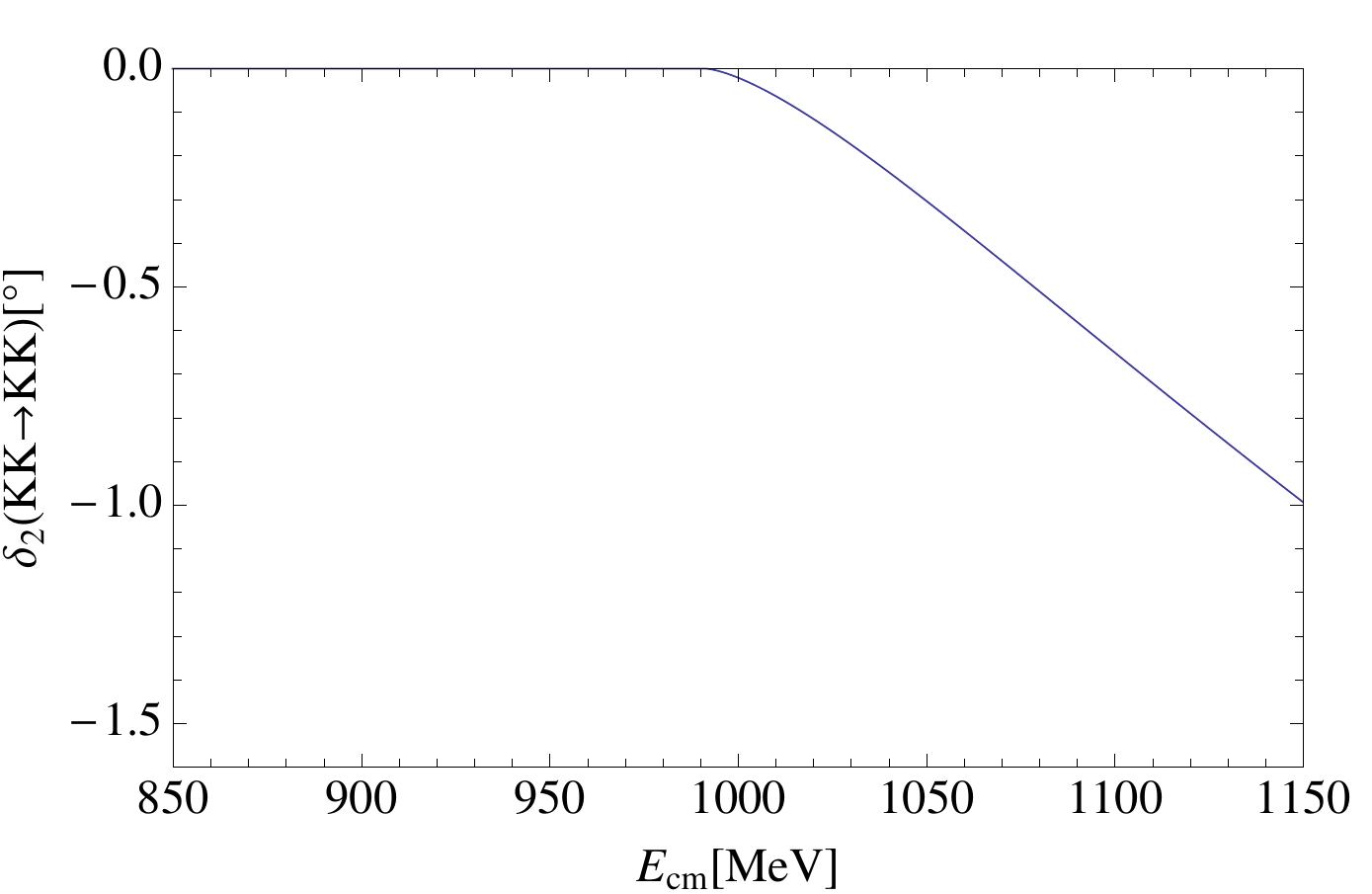}
\end{tabular}
\end{center}
\caption{Left: Elasticity of $\pi\pi\to\pi\pi$ at physical pion masses compared with experimental determinations~\cite{Protopopescu:1973sh}. 
The dashed line shows the inelasticity due to the $K\bar{K}$ channel alone as derived in reference~\cite{Niecknig:2012sj} 
from the Roy-Steiner solution in reference~\cite{Buettiker:2003pp}. 
Right: Phase shift $\delta_2(K\bar{K}\to K\bar{K})$. In this figure we only show the result of variant 2 discussed in the text (results for variant 1 are very similar).}
\label{fig:dkk}
\end{figure*}

The elasticity is shown in Fig. \ref{fig:dkk} (left). It is close to the unity when the $K\bar{K}$ channel is open, and is also consistent with the experimental data and Roy-Steiner equations. The $K\bar{K}$-phase shift is small and negative, as shown in Fig.~\ref{fig:dkk} (right). 
 It has the same sign as determined in Ref.~\cite{Wilson:2015dqa} 
at an unphysical pion mass.

\begin{figure}[ht!]
\begin{center}
\includegraphics[width=0.65\linewidth]{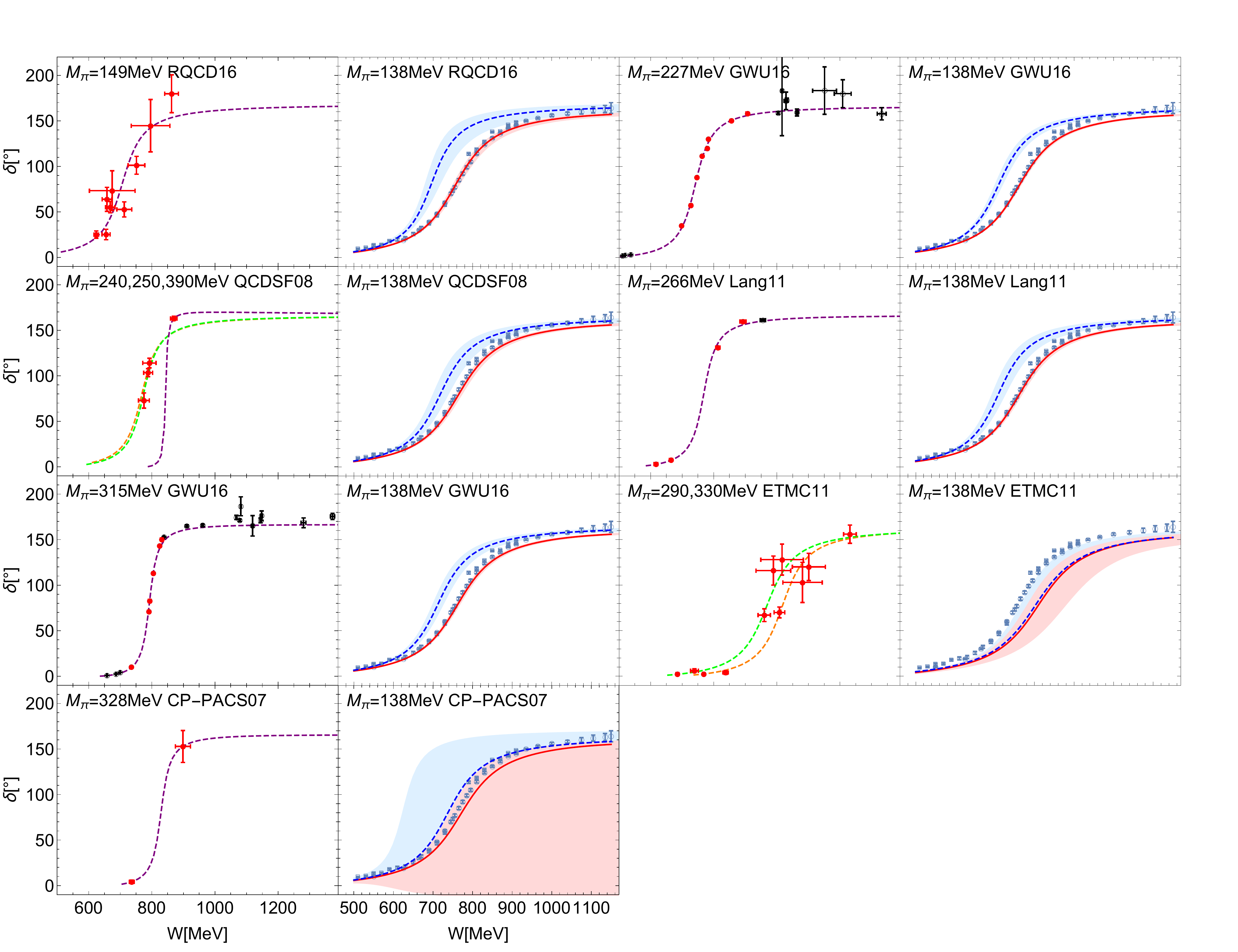}
\end{center}
\caption{Results for the $N_f=2$ lattice simulations (ordered by pion mass) of  Bali {\it et
al.}/RQCD~\cite{Bali:2015gji},  Guo {\it et al.}/GWU~\cite{Guo:2016zos},  G\"ockeler
{\it et al.}/QCDSF~\cite{Gockeler:2008kc}, Lang {\it et
al.}~\cite{Lang:2011mn},  Feng {\it et al.}/ETMC~\cite{Feng:2010es},  Aoki {\it
et al.}/CP-PACS~\cite{Aoki:2007rd}. For each result, the left picture shows the
lattice data and fit, the right figure shows the $N_f=2$ chiral extrapolation
(blue dashed line/light blue area). Without changing this result, the $K\bar K$
channel is then included to predict the effect from the missing strange quark
(red solid line/light read area).  Experimental data (blue circles
from~\cite{Estabrooks:1974vu}, squares from  \cite{Protopopescu:1973sh}) are
then post-dicted. For inherent model uncertainties, see text. }
\label{fig:results1}
\end{figure}

In Fig. \ref{fig:results1} we show the results from fitting the $N_f=2$ lattice data from simulations of the RQCD, GWU ($m_\pi=227$ MeV), QCDSF, Lang {\it et al.}, GWU ($m_\pi=315$ MeV), ETMC and CP-PACS Collaborations, see Refs. \cite{Bali:2015gji,Guo:2016zos,Gockeler:2008kc,Lang:2011mn,Feng:2010es,Aoki:2007rd}, respectively. 
The $68$ \% confidence ellipses in $\hat{l}_1$ and $\hat{l}_2$ all have a common overlap, as shown in the supplemental material of Ref. \cite{Hu:2016shf}. The ellipse from QCDSF Collaboration is very slightly off, while the one 
from ETMC Collaboration is clearly incompatible. Since the uncertainties in the PACS-CS and ETMC analysis are very large we will exclude them in the following discussion. 

The extrapolated results for the phase shifts to the physical mass and the estimated curves when including the $K\bar{K}$ channel are depicted in the right panel of Fig. \ref{fig:results1}. In all the cases (except for the ETMC data) the extrapolation to the physical pion mass is below the experimental data. Switching on the $K\bar{K}$ channel shows significant effects, increasing the $\rho$-mass and leading to a much better prediction. 
To translate the results to the commonly used notation, all phase shifts obtained with the U$\chi$PT model are fitted subsequently 
with the usual Breit-Wigner (BW) parameterization in terms of $g$ and $m_\rho$
(see, e.g., Ref.~\cite{Guo:2016zos}). 

\begin{figure}[ht!]
\begin{center}
\includegraphics[width=0.7\linewidth]{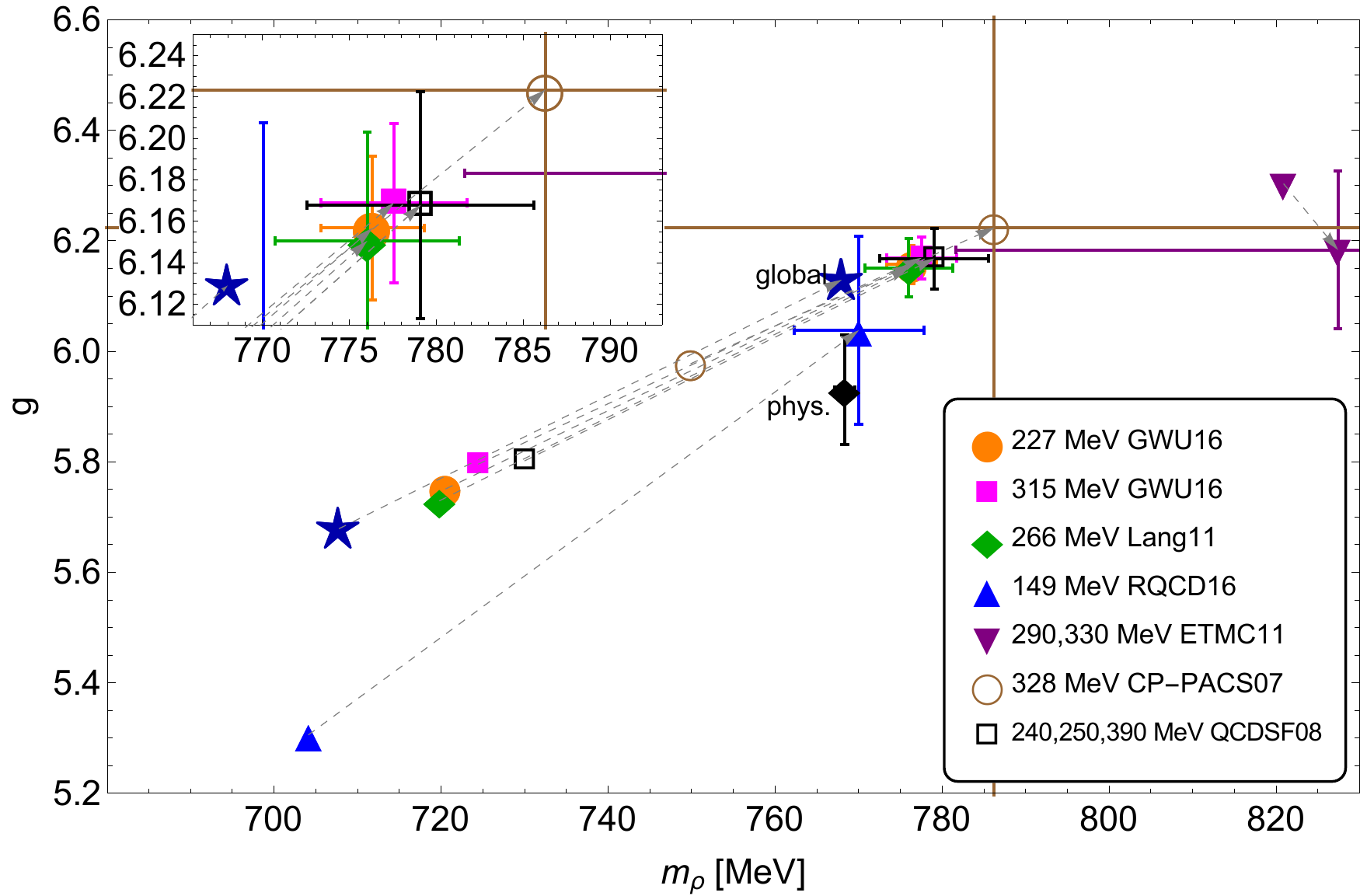}
\end{center}
\caption{Effect of the $K\bar K$ channel in the $(m_\rho,g)$ plane indicated
with arrows, after chiral extrapolation to the physical pion mass.  See
Fig.~\ref{fig:results1} for the labeling of the extrapolations. Only statistical
uncertainties are shown, and only for the case after including $K\bar K$. See text for further
explanations.
}
\label{fig:results2}
\end{figure}

In Fig.~\ref{fig:results2} we show the effect of the $K\bar K$ channel in the
$(m_\rho,g)$ plane. Since $(m_\rho,g)$ emerge from Breit-Wigner fits to
the UCHPT solutions, the comparability with other values in the literature is limited. The experimental point is indicated as ``phys''. In this figure, ``star'' stands for a global fit to the experimental $\pi\pi$ and $\pi K$ phase shift in different isospin
and angular momentum as in Ref. \cite{Doring:2013wka}. It is instructive to remove here the $K\bar K$
channel. As Fig.~\ref{fig:results2} shows (star at $m_\rho\approx 710$~MeV), the
result exhibits the same
trend as the $N_f=2$ lattice data, i.e., a lighter and narrower $\rho$.

The uncertainties in $(m_\rho,g)$ (shown as the error bars in Fig. \ref{fig:results2}) are evaluated as the blue bands in Figs. 
\ref{fig:extrapolation} and \ref{fig:fitfincb} and explained around these figures. 
Once the $K\bar K$ channel is switched on, Fig.~\ref{fig:results2} shows that
$g$ and $m_\rho$ are slightly over-extrapolated. This could be related to model
deficiences. On one side, NLO
contact terms are considered~\cite{Oller:1998hw}, but not the one-loop contributions at NLO as
in Ref.~\cite{Nebreda:2010wv}. On the other side, the LECs entering the
$\pi\pi\to K\bar K$ and $K\bar K\to K\bar K$ transitions are not fully
determined from the fit of lattice data, but from a global fit to $\pi\pi$ and $\pi K$
phase shifts that compromises between the different data sets, leading to a slightly wider $\rho$ resonance. In any case, the estimated
values for the $\rho$ mass are much closer to its physical value after the strange quark is included in all cases except for the CP-PACS and ETMC data analyses, which present larger uncertainties.
\section{Conclusions}
We have performed an analysis of all the available $N_f=2$ lattice data using a U$\chi$PT model. The U$\chi$PT model is able to describe well most of the data sets, with a common overlap of the error ellipses. The extrapolations to the physical pion mass differ significantly from the physical $\rho$ mass. In this talk we have shown that indeed the coupling of the $\rho$ resonance to $K\bar{K}$ can accommodate the observed discrepancies between the $N_f=2$ and $N_f=2+1$ lattice data, leading to an appreciable shift in the $\rho$-mass due to the presence of the strange quark.

\section*{Acknowledgements} 
M. D. gratefully acknowledges support from the NSF/PIF award no. 1415459, the
NSF/Career award no. 1452055. M. D. is also supported by the U.S. Department of Energy, Office of Science, Office of Nuclear Physics under Contract No. DE-AC05-06OR23177. D.G. and A.A. are supported in part by the National Science Foundation CAREER grant PHY-1151648 and Department of Energy grant DE-FG02-95ER40907. The computations were carried out on GWU Colonial One computer cluster and GWU IMPACT collaboration clusters; we are grateful for their support.

\end{document}